\documentclass[11pt]{article}
\usepackage{amsmath, amssymb, amsthm}
\usepackage[hypertexnames=false]{hyperref}
\usepackage{url}
\usepackage{booktabs}
\usepackage{graphicx}
\usepackage{geometry}
\usepackage{indentfirst}
\usepackage{pgfplots}
\usepackage{xcolor}
\pgfplotsset{compat=1.18}
\title{Independent Trivariate Bicycle Codes}
\author{Aygul Azatovna Galimova\textsuperscript{1}\\[4pt]
\textsuperscript{1}Duke University, Durham, NC 27708, USA}
\date{}

\begin{document}
\maketitle

\begin{abstract}
	We introduce six independent trivariate bicycle (ITB) codes, which extend the bivariate bicycle framework of Bravyi et al.\ to three cyclic dimensions. Using asymmetric polynomial pairs on three-dimensional tori, we construct four codes including a $[[140,6,14]]$ code with $kd^2/n = 8.40$. In the code-capacity setting, the $[[140,6,14]]$ code achieves a pseudothreshold of $8.0\%$ and $kd^2/n = 8.40$, exceeding the best multivariate bicycle code of Voss et al.\ ($7.9\%$, $kd^2/n = 2.67$). With circuit-level depolarizing noise, pseudothresholds reach $0.59\%$ for $[[140,6,14]]$ and $0.53\%$ for $[[84,6,10]]$. On the SI1000 superconducting noise model, the $[[140,6,14]]$ code achieves a per-round per-observable rate of $5.6 \times 10^{-5}$ at $p = 0.20\%$. We additionally present two self-dual codes with weight-8 stabilizers: $[[54,14,5]]$ ($kd^2/n = 6.48$) and $[[128,20,8]]$ ($kd^2/n = 10.0$). These results expand the design space of algebraic quantum LDPC codes and demonstrate that the third cyclic dimension yields competitive candidates for practical fault-tolerant implementations.
\end{abstract}

\tableofcontents

\section{Introduction}

\subsection{The Quantum Error Correction Challenge}

	Fault-tolerant quantum computation requires encoding logical quantum information into larger systems of physical qubits. Quantum error-correcting codes protect this information, and the goal is to find codes with good scaling properties: high encoding rates $k/n$ (the ratio of logical to physical qubits), large code distances $d$ (which determine the number of correctable errors), practical stabilizer weights $w$ (which determine measurement complexity), and high error thresholds $p_{\text{th}}$ (the maximum tolerable physical error rate). We say a code has \emph{stabilizer weight}~$w$ if each stabilizer generator acts on at most $w$ qubits.

	One example of a quantum error-correcting code is the surface code~\cite{kitaev2003fault,dennis2002topological,bombin2007optimal}. The surface code achieves a threshold of approximately 1\% under circuit-level depolarizing noise~\cite{fowler2012surface,stephens2014fault}. However, the surface code encodes only a single logical qubit per patch, with code parameters $[[O(d^2), 1, d]]$ and stabilizer weight~4. For example, the rotated surface code~\cite{bombin2007optimal} at distance $d = 5$ requires $25$ physical qubits to protect a single logical qubit.

	The surface code is itself a quantum LDPC code, but with a vanishing encoding rate: $k/n \to 0$ as $n$ grows. Codes with asymptotically non-vanishing encoding rates, where $k/n$ approaches a positive constant, can reduce this overhead~\cite{panteleev2022asymptotically,leverrier2022quantum}. Panteleev and Kalachev~\cite{panteleev2022asymptotically} proved the existence of asymptotically good quantum LDPC codes with constant rate and linear distance; Leverrier and Z\'{e}mor~\cite{leverrier2022quantum} gave a simplified construction with improved distance bounds.

\subsection{Bicycle Codes and Their Extensions}

	The bicycle code construction, introduced by MacKay, Mitchison, and McFadden~\cite{mackay2004sparse}, builds quantum LDPC codes from circulant matrices over a single cyclic shift matrix $S_n$. Panteleev and Kalachev~\cite{panteleev2021degenerate} generalized this construction by allowing two different blocks for the left and right parts of the parity-check matrices. A major advance came with bivariate bicycle (BB) codes~\cite{bravyi2024high}, which use two cyclic shift matrices on a two-dimensional torus $\mathbb{Z}_\ell \times \mathbb{Z}_m$. By introducing long-range connections beyond the nearest-neighbor layout of the surface code, BB codes encode more logical qubits for the same number of physical qubits and code distance. The metric $kd^2/n$ captures this advantage: the surface code has $kd^2/n = 1$, while the BB $[[144,12,12]]$ code achieves $kd^2/n = 12$~\cite{bravyi2024high} with a circuit-level pseudothreshold of $0.65\%$. Yoder et al.~\cite{yoder2025tour} estimate an order-of-magnitude resource advantage over surface codes for BB $[[144,12,12]]$, and Tham et al.~\cite{tham2025modular} demonstrate a logical error rate below $2 \times 10^{-6}$ at physical error rate $10^{-3}$ on trapped-ion hardware.

	Several recent works have sought to expand the algebraic design space beyond the original bivariate construction. These fall into two categories. The first stays within the length-2 chain complex (two blocks in the parity-check matrix): multivariate bicycle codes with three generators subject to $z = xy$~\cite{voss2024multivariate}, generalized toric codes on twisted tori~\cite{liang2025generalized} that achieve $kd^2/n$ ratios exceeding those of BB codes, and coprime bivariate codes~\cite{wang2026coprime} that reach $kd^2/n = 9.97$ with layouts tailored for cold-atom arrays. The second uses longer chain complexes (three or more blocks): trivariate tricycle codes~\cite{jacob2025trivariate} enable partial single-shot decoding and transversal gates, abelian multi-cycle codes~\cite{lin2025abelian} achieve a circuit-level pseudothreshold of approximately $1.1\%$ with weight-6 stabilizers, and multivariate multicycle codes~\cite{mian2026multivariate} provide metachecks for complete single-shot decoding. The longer-chain constructions gain structural properties but increase block length: Jacob et al.'s $[[432,12,12]]$ code requires $3\times$ the qubits of BB $[[144,12,12]]$ at the same $k$ and $d$.

	Our work stays within the length-2 framework but adds a third independent cyclic shift matrix. The length-2 structure keeps the parity-check matrix in the standard $H_X = (A \mid B)$ form, which is compatible with existing syndrome extraction circuits and decoders developed for bivariate bicycle codes. At the same time, the third cyclic dimension increases the polynomial search space from two degrees of freedom to three, enabling codes with higher distances at comparable block lengths. Concretely, our weight-6 codes reach distance $d = 14$ at $n = 140$, whereas bivariate bicycle codes require $n = 144$ to achieve $d = 12$ at the same stabilizer weight.

\subsection{Our Contribution}

	In this work, we introduce independent trivariate bicycle (ITB) codes. These codes extend the bivariate bicycle framework of~\cite{bravyi2024high} to three cyclic dimensions. The two polynomials $A$ and $B$ defining the parity-check matrices are chosen over a three-dimensional torus. We study two variants: \emph{asymmetric} codes, where $A$ and $B$ are chosen independently ($B \neq A^T$), and \emph{self-dual} codes, where $B = A^T$. The asymmetric construction enables weight-6 codes with higher distances than those available at comparable block lengths with the bivariate construction. The self-dual construction yields weight-8 codes that achieve high encoding rates.

	Our main results are as follows:
\begin{enumerate}
\item We construct a family of weight-6 asymmetric codes with $k = 6$ logical qubits (Section~\ref{sec:codes}). The $[[140,6,14]]$ code achieves $kd^2/n = 8.40$.
\item We construct self-dual codes ($B = A^T$) with weight-8 stabilizers and $A = 1 + x + y + z$ (Sections~\ref{sec:self-dual-construction} and~\ref{sec:codes}). The family includes $[[54,14,5]]$ ($kd^2/n = 6.48$) and $[[128,20,8]]$ ($kd^2/n = 10.0$).
\item Under code-capacity noise, the $[[140,6,14]]$ code achieves pseudothreshold $p_0 = 8.0\%$, and the asymmetric codes achieve $kd^2/n$ ratios up to $8.40$ (Section~\ref{sec:voss-comparison}).
\item Under circuit-level depolarizing noise, pseudothresholds reach $0.53\%$ for $[[84,6,10]]$ and $0.59\%$ for $[[140,6,14]]$, with extrapolated logical error rates of $2 \times 10^{-6}$ and $4 \times 10^{-8}$ respectively at $p = 10^{-3}$ (Section~\ref{sec:circuit-level-depol}).
\item Under the SI1000 superconducting noise model~\cite{gidney2022fault} with the Tesseract decoder~\cite{lee2025tesseract}, the $[[140,6,14]]$ code achieves per-round per-observable rate $R_{\mathrm{rnd}} = 5.6 \times 10^{-5}$ at $p = 0.20\%$ (Section~\ref{sec:si1000}).
\end{enumerate}

\section{Mathematical Background}
\label{sec:background}

\subsection{Stabilizer codes and CSS construction}

	A stabilizer code~\cite{gottesman1996class} is defined by an abelian subgroup $\mathcal{S}$ of the $n$-qubit Pauli group that does not contain $-I$. An $[[n, k, d]]$ stabilizer code encodes $k$ logical qubits into $n$ physical qubits with distance $d$.

	Calderbank-Shor-Steane (CSS) codes~\cite{calderbank1996good,steane1996multiple} form a subclass where the stabilizer group separates into $X$-type and $Z$-type generators, captured by parity-check matrices $H_X$ and $H_Z$ satisfying
\begin{equation}
H_X H_Z^T = 0 \pmod{2}.
\label{eq:css-orthogonality}
\end{equation}

	The $X$-distance is $d_X = \min\{\mathrm{wt}(v) : v \in \ker(H_Z) \setminus \mathrm{rowspace}(H_X)\}$, the $Z$-distance is $d_Z = \min\{\mathrm{wt}(v) : v \in \ker(H_X) \setminus \mathrm{rowspace}(H_Z)\}$, and the code distance is $d = \min(d_X, d_Z)$.

\section{Code construction}
\label{sec:construction}

	We follow the notation of~\cite{bravyi2024high}, extended to three generators as in~\cite{jacob2025trivariate}. Let $I_\ell$ and $S_\ell$ be the identity matrix and the cyclic shift matrix of size $\ell \times \ell$ respectively, where the $i$-th row of $S_\ell$ has a single nonzero entry at column $i+1 \pmod{\ell}$. Define the group
\begin{equation}
G = \mathbb{Z}_{\ell_1} \times \mathbb{Z}_{\ell_2} \times \mathbb{Z}_{\ell_3},
\end{equation}
with $N = |G| = \ell_1\ell_2\ell_3$, and the $N \times N$ matrices
\begin{equation}
x = S_{\ell_1} \otimes I_{\ell_2} \otimes I_{\ell_3}, \quad
y = I_{\ell_1} \otimes S_{\ell_2} \otimes I_{\ell_3}, \quad
z = I_{\ell_1} \otimes I_{\ell_2} \otimes S_{\ell_3}.
\end{equation}
Note that $xy = yx$, $xz = zx$, $yz = zy$, and $x^{\ell_1} = y^{\ell_2} = z^{\ell_3} = I_N$. Each monomial $x^a y^b z^c$ is a permutation matrix representing the group element $(a, b, c) \in G$; we use the same symbols for group elements and their matrix representations. Choose $A,B \in \mathbb{F}_2[G]$, written as
\begin{equation}
A = \sum_{j=1}^{\mathcal{W}_A} A_j \quad \text{and} \quad B = \sum_{j=1}^{\mathcal{W}_B} B_j,
\label{eq:AB}
\end{equation}
where each $A_j$ and $B_j$ is a monomial $x^a y^b z^c$. The \emph{weight} of a polynomial is its number of monomial terms, so $A$ has weight $\mathcal{W}_A$ and $B$ has weight $\mathcal{W}_B$. Each row of $H_X$ acts on $\mathcal{W} = \mathcal{W}_A + \mathcal{W}_B$ qubits, so the code has stabilizer weight~$\mathcal{W}$. All matrix arithmetic is over $\mathbb{F}_2$. The pair $(A,B)$ defines a CSS code with check matrices
\begin{equation}
H_X = \begin{pmatrix} A & B \end{pmatrix}, \quad H_Z = \begin{pmatrix} B^T & A^T \end{pmatrix}.
\label{eq:bicycle-parity}
\end{equation}
Both $H_X$ and $H_Z$ have size $N \times 2N$ with block length $n=2N$. The CSS orthogonality condition $H_X H_Z^T = AB^T + BA^T = 0 \pmod{2}$ holds because the group algebra $\mathbb{F}_2[G]$ is commutative for abelian $G$. When $B = A^T$, the code reduces to the \emph{bicycle} construction~\cite{mackay2004sparse} with $H_X = H_Z = (A \mid A^T)$\label{sec:self-dual-construction} and $d_X = d_Z$.

	The number of logical qubits is $k = n - \mathrm{rank}(H_X) - \mathrm{rank}(H_Z)$, where ranks are computed over $\mathbb{F}_2$. The code distance is $d = \min(d_X, d_Z)$, as defined in Section~\ref{sec:background}. All distances reported in this work were computed via integer linear programming (ILP) with branch-and-cut. Five of the six asymmetric codes were independently cross-checked using Stim's weighted maxSAT distance computation~\cite{gidney2021stim}; the sixth ($[[140,6,14]]$) exceeded the solver's time limit. All six codes were additionally verified using the randomized information-set algorithm implemented in the GAP package QDistRnd~\cite{pryadko2022qdistrand}; the upper bounds returned by this algorithm matched the ILP distances in every case.

\section{Code instances}
\label{sec:codes}

	We performed an exhaustive search over all pairs of weight-3 polynomials $A, B \in \mathbb{F}_2[G]$ with $B \neq A^T$ on tori $G = \mathbb{Z}_{\ell_1} \times \mathbb{Z}_{\ell_2} \times \mathbb{Z}_{\ell_3}$ with $\ell_i \leq 7$ and $n = 2\ell_1\ell_2\ell_3 \leq 200$. For each torus, we enumerated all $\binom{N}{3}^2$ polynomial pairs (where $N = \ell_1\ell_2\ell_3$), computed $k$ over $\mathbb{F}_2$, and retained codes with $k > 0$ and $d \geq 4$. Distances were verified as described in Section~\ref{sec:construction}. The $k = 6$ codes achieve the highest $kd^2/n$ ratios among the weight-6 codes found: the $[[140,6,14]]$ code reaches $kd^2/n = 8.40$.

	In a survey of over 170 tori, we observe empirically that $k = 6$ with weight-3 polynomials occurs only when $7 \mid \mathrm{lcm}(\ell_1, \ell_2, \ell_3)$. A partial algebraic explanation is as follows. Let $\mathrm{ord}_2(m)$ denote the multiplicative order of $2$ modulo $m$ (the smallest positive integer $r$ such that $2^r \equiv 1 \pmod{m}$). Since $\mathrm{ord}_2(7) = 3$, the polynomial $x^7 - 1$ factors over $\mathbb{F}_2$ as $(x-1)(x^3 + x + 1)(x^3 + x^2 + 1)$. This factorization gives three nontrivial irreducible $\mathbb{F}_2[\mathbb{Z}_7]$-submodules (of dimensions~1, 3, and~3). These three submodules provide enough independent structure for two weight-3 polynomials to achieve a six-dimensional kernel intersection. Since $\mathrm{ord}_2(p) = 3$ implies $p \mid 2^3 - 1 = 7$, the prime $p = 7$ is the only one with this property. What remains unproven is whether other algebraic mechanisms (involving primes $p$ with $\mathrm{ord}_2(p) \neq 3$) can also yield $k = 6$.

	The self-dual codes use the construction of Section~\ref{sec:self-dual-construction} with $A = 1 + x + y + z$ on cubic tori $\mathbb{Z}_\ell^3$. All codes are listed in Table~\ref{tab:all-codes}.

\begin{table}[htbp]
\centering
\caption{Trivariate codes presented in this work. All distances are ILP-verified. BB codes are listed for comparison.}
\label{tab:all-codes}
\resizebox{\textwidth}{!}{%
\begin{tabular}{lllllcccl}
\toprule
\textbf{Code} & \textbf{Torus} & $A(x,y,z)$ & $B(x,y,z)$ & $k$ & $d$ & \textbf{Wt} & $kd^2/n$ & \textbf{Ref.} \\
\midrule
\multicolumn{9}{l}{\emph{BB reference codes}} \\
BB $[[72,12,6]]$ & $6 \times 6$ & $x^3+y+y^2$ & $y^3+x+x^2$ & 12 & 6 & 6 & 6.00 & \cite{bravyi2024high} \\
BB $[[144,12,12]]$ & $12 \times 6$ & $x^3+y+y^2$ & $y^3+x+x^2$ & 12 & 12 & 6 & 12.00 & \cite{bravyi2024high} \\
\midrule
\multicolumn{9}{l}{\emph{$k=6$ asymmetric codes}} \\
$[[84,6,10]]$ & $2 \times 3 \times 7$ & $1+y^2z^4+xyz^5$ & $1+z+xyz^3$ & 6 & 10 & 6 & 7.14 & this work \\
$[[140,6,14]]$ & $2 \times 5 \times 7$ & $1+yz^3+xyz^2$ & $1+xy^4z^2+xy^4z^3$ & 6 & 14 & 6 & 8.40 & this work \\
$[[196,6,12]]$ & $2 \times 7 \times 7$ & $1+xz^2+xy^3z^6$ & $1+xyz^6+xy^3z^2$ & 6 & 12 & 6 & 4.41 & this work \\
\midrule
\multicolumn{9}{l}{\emph{Other weight-6 asymmetric codes}} \\
$[[54,8,6]]$ & $3 \times 3 \times 3$ & $1+z^2+xz$ & $1+xy+xy^2$ & 8 & 6 & 6 & 5.33 & this work \\
\midrule
\multicolumn{9}{l}{\emph{Self-dual codes ($B = A^T$)}} \\
$[[54,14,5]]$ & $3 \times 3 \times 3$ & $1+x+y+z$ & $1+x^2+y^2+z^2$ & 14 & 5 & 8 & 6.48 & this work \\
$[[128,20,8]]$ & $4 \times 4 \times 4$ & $1+x+y+z$ & $1+x^3+y^3+z^3$ & 20 & 8 & 8 & 10.0 & this work \\
\bottomrule
\end{tabular}%
}
\end{table}

\section{Noise simulations}
\label{sec:simulations}

	We evaluate our codes under three noise models: code-capacity, circuit-level depolarizing (decoded with BP-OSD from the LDPC library~\cite{roffe2020ldpc}), and the SI1000 superconducting noise model~\cite{gidney2022fault} (decoded with Tesseract~\cite{lee2025tesseract}).

\subsection{Code-capacity noise results}
\label{sec:code-capacity}

	Under code-capacity noise, each data qubit suffers independent depolarizing noise at rate $p$ while syndrome measurements are perfect (no measurement errors, no syndrome extraction rounds). The decoder is BP-OSD~\cite{roffe2020ldpc} (min-sum, $\texttt{max\_iter}=50$, $\texttt{osd\_order}=10$); this matches the decoder class of Voss et al.~\cite{voss2024multivariate}. Physical error rates range from $p = 10^{-3}$ to $p \approx 0.12$. The pseudothreshold $p_0$ is the physical error rate at which the logical error rate crosses $p_L = p$. Tables~\ref{tab:vs-bb} and~\ref{tab:vs-voss} report results; Section~\ref{sec:voss-comparison} discusses the comparison.

\subsection{Circuit-level depolarizing noise results}
\label{sec:circuit-level-depol}

	To enable direct comparison with Bravyi et al.~\cite{bravyi2024high}, we use their exact single-fault enumeration decoder under uniform circuit-level depolarizing noise: each CNOT gate fails with probability $p$ (one of 15 nontrivial two-qubit Paulis, each with probability $p/15$), idle qubits suffer depolarizing noise at rate $p$, state preparation and measurement each fail with probability $p$. Each trial runs $N_c = 12$ syndrome cycles for all codes, decoded with BP-OSD (min-sum, $10{,}000$ iterations, OSD-CS order 7). This matches the decoder, noise model, and cycle count of the reference implementation of~\cite{bravyi2024high}.

	Table~\ref{tab:depol-results} reports pseudothresholds and extrapolated logical error rates. Following~\cite{bravyi2024high}, a trial is counted as failed if any logical operator is corrupted; let $P_{\mathrm{any}}$ denote the fraction of failed trials. The per-round logical error rate is $p_L = 1-(1-P_{\mathrm{any}})^{1/N_c}$. The pseudothreshold $p_0$ solves the break-even equation $p_L(p_0) = k \cdot p_0$. Both $p_0$ and the extrapolated $p_L$ values are obtained by fitting $\log p_L - (d/2)\log p = c_0 + c_1 p + c_2 p^2$, where $c_0$, $c_1$, $c_2$ are fitting parameters, to Monte Carlo data at $p \ge 2 \times 10^{-3}$. This fitting range matches that of~\cite{bravyi2024high} for BB $[[72,12,6]]$. The BB pseudothresholds are from~\cite{bravyi2024high} (Table~1); all $p_L$ values (including BB) are from our Monte Carlo data.

\begin{table}[htbp]
\centering
\caption{Performance under the circuit-level depolarizing noise model of~\cite{bravyi2024high}. The net encoding rate is $r = \lfloor 2n/k \rfloor^{-1}$, accounting for $n$ ancilla qubits. Each trial runs $N_c = 12$ syndrome cycles, decoded with BP-OSD (min-sum, $10{,}000$ iterations, OSD-CS order~7) as in the reference implementation of~\cite{bravyi2024high}. The per-round logical error rate $p_L = 1-(1-P_{\mathrm{any}})^{1/N_c}$ is computed from $10{,}000$--$60{,}000$ Monte Carlo trials per point and extrapolated via $p_L = p^{d/2}\exp(c_0 + c_1 p + c_2 p^2)$. All codes are fit at $p \ge 2 \times 10^{-3}$, the same fitting range as~\cite{bravyi2024high}. The pseudothreshold $p_0$ solves $p_L(p_0) = k \cdot p_0$. BB pseudothresholds are from~\cite{bravyi2024high} (Table~1); BB $p_L$ values are from our Monte Carlo data.}
\label{tab:depol-results}
\small
\begin{tabular}{lccccccc}
\toprule
\textbf{Code} & $d$ & $r$ & $kd^2/n$ & $p_0$ & $p_L(10^{-3})$ & $p_L(10^{-4})$ \\
\midrule
BB $[[144,12,12]]$ & $12$ & $1/24$ & $12.00$ & $0.65\%$ & $2 \times 10^{-7}$ & $7 \times 10^{-14}$ \\
BB $[[72,12,6]]$ & $6$ & $1/12$ & $6.00$ & $0.48\%$ & $7 \times 10^{-5}$ & $4 \times 10^{-8}$ \\
\midrule
$[[140,6,14]]$ & $14$ & $1/46$ & $8.40$ & $0.59\%$ & $4 \times 10^{-8}$ & $3 \times 10^{-15}$ \\
$[[196,6,12]]$ & $12$ & $1/65$ & $4.41$ & $0.60\%$ & $7 \times 10^{-8}$ & $2 \times 10^{-14}$ \\
$[[84,6,10]]$ & $10$ & $1/28$ & $7.14$ & $0.53\%$ & $2 \times 10^{-6}$ & $8 \times 10^{-12}$ \\
$[[128,20,8]]$ & $8$ & $1/12$ & $10.00$ & $0.45\%$ & $2 \times 10^{-5}$ & $7 \times 10^{-10}$ \\
$[[54,14,5]]$ & $5$ & $1/7$ & $6.48$ & $0.33\%$ & $1 \times 10^{-3}$ & $2 \times 10^{-6}$ \\
$[[54,8,6]]$ & $6$ & $1/13$ & $5.33$ & $0.48\%$ & $1 \times 10^{-4}$ & $8 \times 10^{-8}$ \\
\bottomrule
\end{tabular}
\end{table}

\begin{figure}[htbp]
\centering
\definecolor{clrBBa}{HTML}{D62728}%
\definecolor{clrBBb}{HTML}{1F77B4}%
\definecolor{clrA}{HTML}{FF7F0E}%
\definecolor{clrB}{HTML}{2CA02C}%
\definecolor{clrC}{HTML}{9467BD}%
\definecolor{clrE}{HTML}{E377C2}%
\begin{tikzpicture}
\begin{loglogaxis}[
    width=14cm, height=10cm,
    xmin=8e-4, xmax=1e-2, ymin=1e-9, ymax=1e-1,
    xlabel={Physical error rate $p$},
    ylabel={Per-round logical error rate $p_L$},
    grid=both,
    major grid style={line width=0.3pt, draw=black!20},
    minor grid style={line width=0.2pt, draw=black!8},
    legend pos=south east,
    legend style={font=\footnotesize, fill opacity=0.9, draw=black!20, cells={anchor=west}},
    tick label style={font=\small}, label style={font=\small},
    every error bar/.style={line width=0.4pt},
]
\addplot[clrBBa, thick, no markers] coordinates {
(1.0000e-04,7.4749e-14)(1.1669e-04,1.9149e-13)(1.3618e-04,4.9168e-13)
(1.5891e-04,1.2659e-12)(1.8544e-04,3.2697e-12)(2.1640e-04,8.4759e-12)
(2.5253e-04,2.2064e-11)(2.9468e-04,5.7717e-11)(3.4388e-04,1.5182e-10)
(4.0129e-04,4.0194e-10)(4.6829e-04,1.0719e-09)(5.4647e-04,2.8826e-09)
(6.3770e-04,7.8257e-09)(7.4416e-04,2.1473e-08)(8.6839e-04,5.9627e-08)
(1.0134e-03,1.6778e-07)(1.1825e-03,4.7901e-07)(1.3800e-03,1.3891e-06)
(1.6104e-03,4.0951e-06)(1.8792e-03,1.2278e-05)(2.1929e-03,3.7418e-05)
(2.5590e-03,1.1572e-04)(2.9862e-03,3.6189e-04)(3.4848e-03,1.1370e-03)
(4.0666e-03,3.5521e-03)(4.7455e-03,1.0856e-02)(5.5377e-03,3.1674e-02)
(6.4622e-03,8.5099e-02)(7.5410e-03,1.9981e-01)(8.8000e-03,3.8038e-01)
};\addlegendentry{BB $[[144{,}12{,}12]]$}
\addplot[forget plot, only marks, mark=diamond*, mark size=3pt,
    clrBBa, mark options={draw=black, line width=0.3pt},
    error bars/.cd, y dir=both, y explicit]
coordinates {
  (0.003, 3.8414e-04) +- (0, 1.1213e-04)
  (0.004, 3.1024e-03) +- (0, 3.1773e-04)
  (0.005, 1.6292e-02) +- (0, 7.5005e-04)
  (0.006, 5.5702e-02) +- (0, 1.5340e-03)
  (0.007, 1.3157e-01) +- (0, 2.9866e-03)
};
\addplot[clrBBb, thick, no markers] coordinates {
(1.0000e-04,3.7810e-08)(1.1669e-04,6.0781e-08)(1.3618e-04,9.7892e-08)
(1.5891e-04,1.5801e-07)(1.8544e-04,2.5569e-07)(2.1640e-04,4.1499e-07)
(2.5253e-04,6.7583e-07)(2.9468e-04,1.1050e-06)(3.4388e-04,1.8148e-06)
(4.0129e-04,2.9961e-06)(4.6829e-04,4.9762e-06)(5.4647e-04,8.3216e-06)
(6.3770e-04,1.4025e-05)(7.4416e-04,2.3846e-05)(8.6839e-04,4.0949e-05)
(1.0134e-03,7.1103e-05)(1.1825e-03,1.2499e-04)(1.3800e-03,2.2267e-04)
(1.6104e-03,4.0245e-04)(1.8792e-03,7.3847e-04)(2.1929e-03,1.3760e-03)
(2.5590e-03,2.6019e-03)(2.9862e-03,4.9843e-03)(3.4848e-03,9.6369e-03)
(4.0666e-03,1.8687e-02)(4.7455e-03,3.5967e-02)(5.5377e-03,6.7631e-02)
(6.4622e-03,1.2130e-01)(7.5410e-03,2.0034e-01)(8.8000e-03,2.8958e-01)
};\addlegendentry{BB $[[72{,}12{,}6]]$}
\addplot[forget plot, only marks, mark=diamond*, mark size=3pt,
    clrBBb, mark options={draw=black, line width=0.3pt},
    error bars/.cd, y dir=both, y explicit]
coordinates {
  (0.002, 8.4559e-04) +- (0, 1.6560e-04)
  (0.003, 4.8612e-03) +- (0, 3.9915e-04)
  (0.004, 1.7629e-02) +- (0, 7.8273e-04)
  (0.005, 4.3627e-02) +- (0, 1.3142e-03)
  (0.006, 8.8734e-02) +- (0, 2.1307e-03)
  (0.007, 1.5722e-01) +- (0, 3.5857e-03)
  (0.008, 2.3887e-01) +- (0, 6.2690e-03)
};
\addplot[clrA, thick, dashed, no markers] coordinates {
(1.0000e-04,2.6617e-15)(1.1669e-04,7.9157e-15)(1.3618e-04,2.3576e-14)
(1.5891e-04,7.0339e-14)(1.8544e-04,2.1028e-13)(2.1640e-04,6.3010e-13)
(2.5253e-04,1.8931e-12)(2.9468e-04,5.7055e-12)(3.4388e-04,1.7257e-11)
(4.0129e-04,5.2406e-11)(4.6829e-04,1.5989e-10)(5.4647e-04,4.9042e-10)
(6.3770e-04,1.5132e-09)(7.4416e-04,4.7009e-09)(8.6839e-04,1.4714e-08)
(1.0134e-03,4.6439e-08)(1.1825e-03,1.4790e-07)(1.3800e-03,4.7568e-07)
(1.6104e-03,1.5455e-06)(1.8792e-03,5.0735e-06)(2.1929e-03,1.6820e-05)
(2.5590e-03,5.6240e-05)(2.9862e-03,1.8917e-04)(3.4848e-03,6.3724e-04)
(4.0666e-03,2.1343e-03)(4.7455e-03,7.0291e-03)(5.5377e-03,2.2387e-02)
(6.4622e-03,6.7286e-02)(7.5410e-03,1.8423e-01)(8.8000e-03,4.3690e-01)
};\addlegendentry{$[[140{,}6{,}14]]$}
\addplot[clrB, thick, dashed, no markers] coordinates {
(1.0000e-04,8.0473e-12)(1.1669e-04,1.7645e-11)(1.3618e-04,3.8774e-11)
(1.5891e-04,8.5416e-11)(1.8544e-04,1.8871e-10)(2.1640e-04,4.1829e-10)
(2.5253e-04,9.3079e-10)(2.9468e-04,2.0804e-09)(3.4388e-04,4.6735e-09)
(4.0129e-04,1.0560e-08)(4.6829e-04,2.4021e-08)(5.4647e-04,5.5055e-08)
(6.3770e-04,1.2726e-07)(7.4416e-04,2.9700e-07)(8.6839e-04,7.0056e-07)
(1.0134e-03,1.6719e-06)(1.1825e-03,4.0413e-06)(1.3800e-03,9.9011e-06)
(1.6104e-03,2.4599e-05)(1.8792e-03,6.1971e-05)(2.1929e-03,1.5814e-04)
(2.5590e-03,4.0781e-04)(2.9862e-03,1.0581e-03)(3.4848e-03,2.7418e-03)
(4.0666e-03,7.0135e-03)(4.7455e-03,1.7400e-02)(5.5377e-03,4.0779e-02)
(6.4622e-03,8.6889e-02)(7.5410e-03,1.5928e-01)(8.8000e-03,2.3222e-01)
};\addlegendentry{$[[84{,}6{,}10]]$}
\addplot[clrC, thick, dashed, no markers] coordinates {
(1.0000e-04,4.8516e-14)(1.1669e-04,1.2333e-13)(1.3618e-04,3.1385e-13)
(1.5891e-04,7.9973e-13)(1.8544e-04,2.0410e-12)(2.1640e-04,5.2180e-12)
(2.5253e-04,1.3369e-11)(2.9468e-04,3.4337e-11)(3.4388e-04,8.8452e-11)
(4.0129e-04,2.2864e-10)(4.6829e-04,5.9341e-10)(5.4647e-04,1.5476e-09)
(6.3770e-04,4.0591e-09)(7.4416e-04,1.0719e-08)(8.6839e-04,2.8535e-08)
(1.0134e-03,7.6697e-08)(1.1825e-03,2.0855e-07)(1.3800e-03,5.7498e-07)
(1.6104e-03,1.6121e-06)(1.8792e-03,4.6124e-06)(2.1929e-03,1.3527e-05)
(2.5590e-03,4.0891e-05)(2.9862e-03,1.2830e-04)(3.4848e-03,4.2149e-04)
(4.0666e-03,1.4663e-03)(4.7455e-03,5.4797e-03)(5.5377e-03,2.2410e-02)
(6.4622e-03,1.0273e-01)(7.5410e-03,5.4461e-01)(8.8000e-03,3.4780e+00)
};\addlegendentry{$[[196{,}6{,}12]]$}
\addplot[forget plot, only marks, mark=*, mark size=2pt,
    clrA, mark options={draw=black, line width=0.3pt},
    error bars/.cd, y dir=both, y explicit]
coordinates {
  (0.003, 1.4178e-04) +- (0, 6.9258e-05)
  (0.004, 1.7671e-03) +- (0, 2.3930e-04)
  (0.005, 1.1728e-02) +- (0, 6.2960e-04)
  (0.006, 4.3003e-02) +- (0, 1.3027e-03)
  (0.007, 1.0659e-01) +- (0, 2.4704e-03)
};
\addplot[forget plot, only marks, mark=square*, mark size=2pt,
    clrB, mark options={draw=black, line width=0.3pt},
    error bars/.cd, y dir=both, y explicit]
coordinates {
  (0.003, 1.0814e-03) +- (0, 1.8718e-04)
  (0.004, 6.7716e-03) +- (0, 4.7305e-04)
  (0.005, 2.3040e-02) +- (0, 9.0657e-04)
  (0.006, 5.8632e-02) +- (0, 1.5865e-03)
  (0.007, 1.1768e-01) +- (0, 2.6927e-03)
  (0.008, 1.9801e-01) +- (0, 4.7440e-03)
};
\addplot[forget plot, only marks, mark=triangle*, mark size=2.5pt,
    clrC, mark options={draw=black, line width=0.3pt},
    error bars/.cd, y dir=both, y explicit]
coordinates {
  (0.003, 1.4178e-04) +- (0, 6.9258e-05)
  (0.004, 1.2249e-03) +- (0, 1.9919e-04)
  (0.005, 8.8426e-03) +- (0, 5.4309e-04)
  (0.006, 4.1013e-02) +- (0, 1.2656e-03)
  (0.007, 1.1673e-01) +- (0, 2.6733e-03)
  (0.008, 2.3354e-01) +- (0, 6.0440e-03)
};
\addplot[clrE, thick, dashed, no markers] coordinates {
(1.0000e-04,7.8780e-08)(1.1669e-04,1.2620e-07)(1.3618e-04,2.0243e-07)
(1.5891e-04,3.2522e-07)(1.8544e-04,5.2339e-07)(2.1640e-04,8.4406e-07)
(2.5253e-04,1.3644e-06)(2.9468e-04,2.2117e-06)(3.4388e-04,3.5963e-06)
(4.0129e-04,5.8691e-06)(4.6829e-04,9.6181e-06)(5.4647e-04,1.5837e-05)
(6.3770e-04,2.6217e-05)(7.4416e-04,4.3666e-05)(8.6839e-04,7.3228e-05)
(1.0134e-03,1.2374e-04)(1.1825e-03,2.1087e-04)(1.3800e-03,3.6264e-04)
(1.6104e-03,6.2976e-04)(1.8792e-03,1.1048e-03)(2.1929e-03,1.9578e-03)
(2.5590e-03,3.5025e-03)(2.9862e-03,6.3161e-03)(3.4848e-03,1.1447e-02)
(4.0666e-03,2.0746e-02)(4.7455e-03,3.7301e-02)(5.5377e-03,6.5735e-02)
(6.4622e-03,1.1150e-01)(7.5410e-03,1.7732e-01)(8.8000e-03,2.5446e-01)
};\addlegendentry{$[[54{,}8{,}6]]$}
\addplot[forget plot, only marks, mark=pentagon*, mark size=2.5pt,
    clrE, mark options={draw=black, line width=0.3pt},
    error bars/.cd, y dir=both, y explicit]
coordinates {
  (0.002, 1.5125e-03) +- (0, 2.2136e-04)
  (0.003, 6.4399e-03) +- (0, 4.6098e-04)
  (0.004, 1.9939e-02) +- (0, 8.3705e-04)
  (0.005, 4.5484e-02) +- (0, 1.3485e-03)
  (0.006, 8.4052e-02) +- (0, 2.0444e-03)
  (0.007, 1.3847e-01) +- (0, 3.1399e-03)
  (0.008, 2.0588e-01) +- (0, 5.0048e-03)
};
\end{loglogaxis}
\end{tikzpicture}
\caption{Per-round logical error rate $p_L$ versus physical error rate $p$ under circuit-based depolarizing noise~\cite{bravyi2024high}. Solid lines: BB codes; dashed: ours. Curves are fits $p_L = p^{d/2}\exp(c_0 + c_1 p + c_2 p^2)$; markers: Monte Carlo estimates ($10{,}000$--$60{,}000$ trials, $N_c = 12$ rounds); bars: $95\%$ Wilson intervals. BB data reproduces the published values of~\cite{bravyi2024high} (Table~1).}
\label{fig:circuit-level-depol}
\end{figure}
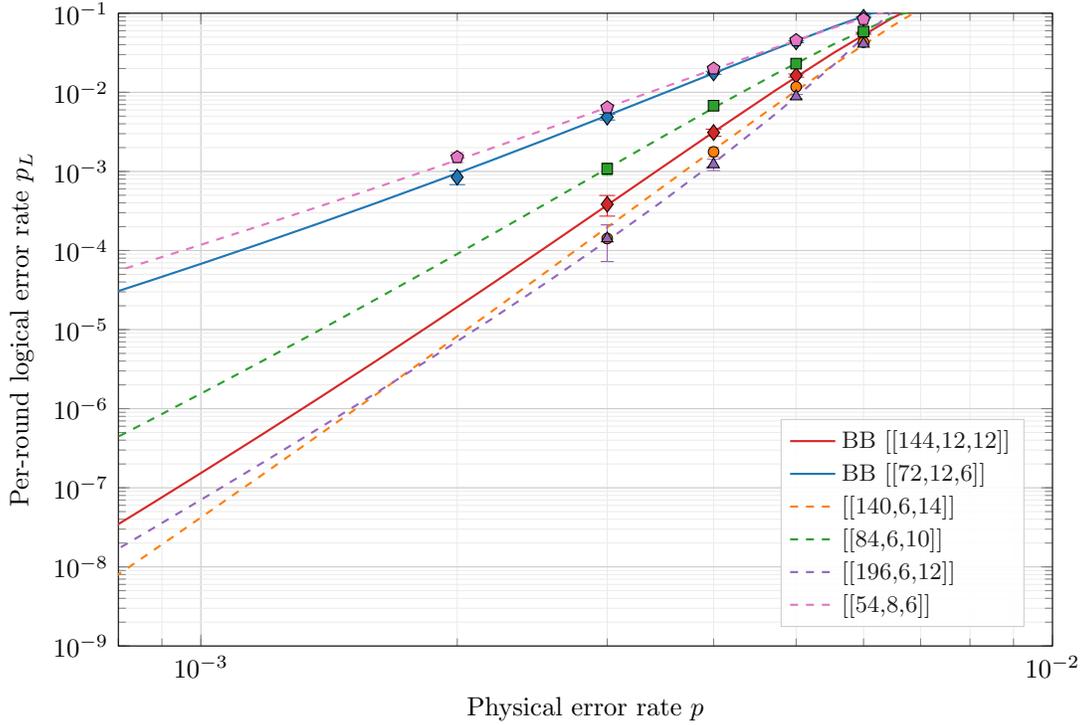

	The $[[84,6,10]]$ code achieves a pseudothreshold of $p_0 \approx 0.53\%$ and the $[[140,6,14]]$ code achieves $p_0 \approx 0.59\%$. At $p = 10^{-3}$, the extrapolated logical error rates are $2 \times 10^{-6}$ for $[[84,6,10]]$ and $4 \times 10^{-8}$ for $[[140,6,14]]$. For reference, BB $[[72,12,6]]$ has $p_0 = 0.48\%$ with $p_L(10^{-3}) = 7 \times 10^{-5}$, and BB $[[144,12,12]]$ has $p_0 = 0.65\%$ with $p_L(10^{-3}) = 2 \times 10^{-7}$; these codes encode $k = 12$ logical qubits compared to $k = 6$ for our asymmetric codes. The self-dual code $[[128,20,8]]$ encodes $k = 20$ logical qubits and achieves $p_0 \approx 0.45\%$ with $kd^2/n = 10.00$. Our BB Monte Carlo data reproduces the published values of~\cite{bravyi2024high} (Table~1). Figure~\ref{fig:circuit-level-depol} shows the error rate curves and their extrapolations.

\subsection{SI1000 noise results}
\label{sec:si1000}

	Table~\ref{tab:si1000-results} shows per-round per-observable error rates under the SI1000 noise model~\cite{gidney2022fault}, decoded using Tesseract~\cite{lee2025tesseract} with $10{,}000$--$400{,}000$ shots per data point. At $p = 0.05\%$, the codes $[[84,6,10]]$, $[[140,6,14]]$, $[[196,6,12]]$, $[[128,20,8]]$, and BB $[[144,12,12]]$ all achieve zero observed errors. At $p = 0.10\%$, the $[[84,6,10]]$ code gives $R_{\mathrm{rnd}} = 1.7 \times 10^{-7}$ ($1$ error in $200{,}000$ shots), while $[[140,6,14]]$, $[[196,6,12]]$, and BB $[[144,12,12]]$ remain at zero. At $p = 0.20\%$, the $[[140,6,14]]$ code gives $R_{\mathrm{rnd}} = 5.6 \times 10^{-5}$, the $[[84,6,10]]$ code gives $R_{\mathrm{rnd}} = 7.6 \times 10^{-5}$, and BB $[[144,12,12]]$ gives $R_{\mathrm{rnd}} = 9.2 \times 10^{-5}$. BB $[[72,12,6]]$ gives $R_{\mathrm{rnd}} = 4.3 \times 10^{-4}$ at the same error rate.

\begin{table}[htbp]
\centering
\caption{Per-round per-observable error rates under the SI1000 noise model~\cite{gidney2022fault} ($d$ syndrome rounds, interleaved schedules, Tesseract decoder~\cite{lee2025tesseract}, $10{,}000$--$400{,}000$ shots). The per-round rate is $R_{\mathrm{rnd}} = \tfrac{1}{2}(1-(1-2R)^{1/d})$, following~\cite{lee2025tesseract}. Entries marked $0$ had no observed errors.}
\label{tab:si1000-results}
\small
\begin{tabular}{llccccc}
\toprule
& & & & \multicolumn{3}{c}{\textbf{Per-round per-observable rate $R_{\mathrm{rnd}}$}} \\
\cmidrule(lr){5-7}
\textbf{Code} & $n$ & $k$ & $d$ & $p{=}0.05\%$ & $p{=}0.10\%$ & $p{=}0.20\%$ \\
\midrule
BB $[[72,12,6]]$ & 72 & 12 & 6 & $4.6 \times 10^{-6}$ & $3.0 \times 10^{-5}$ & $4.3 \times 10^{-4}$ \\
BB $[[144,12,12]]$ & 144 & 12 & 12 & $0$ & $0$ & $9.2 \times 10^{-5}$ \\
\midrule
$[[140,6,14]]$ & 140 & 6 & 14 & $0$ & $0$ & $5.6 \times 10^{-5}$ \\
$[[84,6,10]]$ & 84 & 6 & 10 & $0$ & $1.7 \times 10^{-7}$ & $7.6 \times 10^{-5}$ \\
$[[196,6,12]]$ & 196 & 6 & 12 & $0$ & $0$ & $8.3 \times 10^{-5}$ \\
$[[54,8,6]]$ & 54 & 8 & 6 & $4.6 \times 10^{-6}$ & $3.5 \times 10^{-5}$ & $4.9 \times 10^{-4}$ \\
$[[128,20,8]]$ & 128 & 20 & 8 & $0$ & $1.4 \times 10^{-5}$ & $2.5 \times 10^{-3}$ \\
$[[54,14,5]]$ & 54 & 14 & 5 & $2.8 \times 10^{-5}$ & $1.4 \times 10^{-4}$ & $3.0 \times 10^{-3}$ \\
\bottomrule
\end{tabular}
\end{table}

\section{Comparison with existing codes}
\label{sec:comparison}

\subsection{Code parameters}

	Table~\ref{tab:final-comparison} compares our codes with existing constructions. Among weight-6 codes, the $[[140,6,14]]$ code achieves the highest distance ($d=14$) and $kd^2/n = 8.40$. The weight-8 code $[[128,20,8]]$ reaches $kd^2/n = 10.0$, trading stabilizer weight for encoding rate.

\begin{table}[htbp]
\centering
\caption{Comparison with existing quantum LDPC codes. Asymmetric: $B \neq A^T$; self-dual: $B = A^T$.}
\label{tab:final-comparison}
\small
\begin{tabular}{lcccccl}
\toprule
\textbf{Code} & $n$ & $k$ & $d$ & $kd^2/n$ & \textbf{Wt} & \textbf{Ref.} \\
\midrule
Surface code~\cite{bombin2007optimal} & $d^2$ & 1 & $d$ & $1$ & 4 & \cite{bombin2007optimal} \\
BB $[[144,12,12]]$ & 144 & 12 & 12 & 12.00 & 6 & \cite{bravyi2024high} \\
BB $[[72,12,6]]$ & 72 & 12 & 6 & 6.00 & 6 & \cite{bravyi2024high} \\
Voss $[[96,4,8]]$ & 96 & 4 & 8 & 2.67 & 5 & \cite{voss2024multivariate} \\
Voss $[[48,4,6]]$ & 48 & 4 & 6 & 3.00 & 6 & \cite{voss2024multivariate} \\
Voss $[[30,4,5]]$ & 30 & 4 & 5 & 3.33 & 5 & \cite{voss2024multivariate} \\
Jacob TT $[[72,6,6]]$ & 72 & 6 & 6 & 3.00 & 6--9 & \cite{jacob2025trivariate} \\
Jacob TT $[[180,12,8]]$ & 180 & 12 & 8 & 4.27 & 6--9 & \cite{jacob2025trivariate} \\
Jacob TT $[[432,12,12]]$ & 432 & 12 & 12 & 4.00 & 6--9 & \cite{jacob2025trivariate} \\
Lin 4D toric & var & var & var & -- & 6 & \cite{lin2025abelian} \\
\midrule
$[[196,6,12]]$ & 196 & 6 & 12 & 4.41 & 6 & this work \\
$[[140,6,14]]$ & 140 & 6 & 14 & 8.40 & 6 & this work \\
$[[84,6,10]]$ & 84 & 6 & 10 & 7.14 & 6 & this work \\
$[[54,8,6]]$ & 54 & 8 & 6 & 5.33 & 6 & this work \\
$[[128,20,8]]$ & 128 & 20 & 8 & 10.00 & 8 & this work \\
$[[54,14,5]]$ & 54 & 14 & 5 & 6.48 & 8 & this work \\
\bottomrule
\end{tabular}
\end{table}

\subsection{Code-capacity comparison}
\label{sec:voss-comparison}

	Under code-capacity noise (Section~\ref{sec:code-capacity}), we compare with the multivariate bicycle codes of Voss et al.~\cite{voss2024multivariate} and the BB codes of Bravyi et al.~\cite{bravyi2024high}. Tables~\ref{tab:vs-bb} and~\ref{tab:vs-voss} report pseudothreshold and $kd^2/n$ ratios.

\begin{table}[htbp]
\centering
\caption{Our codes versus Bravyi--Cross BB codes under code-capacity noise (BP-OSD, $\texttt{osd\_order}=10$). Ratios above~1 indicate our code has the higher value.}
\label{tab:vs-bb}
\small
\begin{tabular}{lcccccc}
\toprule
Our code & $kd^2/n$ & $p_0$ & \multicolumn{2}{c}{vs BB $[[72,12,6]]$} & \multicolumn{2}{c}{vs BB $[[144,12,12]]$} \\
\cmidrule(lr){4-5}\cmidrule(lr){6-7}
 & & & $p_0$ ratio & $kd^2/n$ ratio & $p_0$ ratio & $kd^2/n$ ratio \\
\midrule
$[[84,6,10]]$ wt-6  & 7.14 & 7.10\% & $2.01\times$ & $1.19\times$ & $0.98\times$ & $0.60\times$ \\
$[[140,6,14]]$ wt-6 & 8.40 & 8.02\% & $2.27\times$ & $1.40\times$ & $1.11\times$ & $0.70\times$ \\
$[[196,6,12]]$ wt-6 & 4.41 & 8.49\% & $2.40\times$ & $0.74\times$ & $1.18\times$ & $0.37\times$ \\
$[[54,8,6]]$ wt-6   & 5.33 & 3.46\% & $0.98\times$ & $0.89\times$ & $0.48\times$ & $0.44\times$ \\
$[[54,14,5]]$ wt-8  & 6.48 & 2.57\% & $0.73\times$ & $1.08\times$ & $0.36\times$ & $0.54\times$ \\
$[[128,20,8]]$ wt-8 & 10.00 & 5.05\% & $1.43\times$ & $1.67\times$ & $0.70\times$ & $0.83\times$ \\
\bottomrule
\end{tabular}
\end{table}

\begin{table}[htbp]
\centering
\caption{Our codes versus Voss et al.~\cite{voss2024multivariate} under code-capacity noise. Each row pairs our code with the most comparable Voss code by distance or stabilizer weight.}
\label{tab:vs-voss}
\small
\begin{tabular}{llcccc}
\toprule
Our code & Voss et al.\ & \multicolumn{2}{c}{$p_0$} & \multicolumn{2}{c}{$kd^2/n$} \\
\cmidrule(lr){3-4}\cmidrule(lr){5-6}
 & & Ours / Voss & Ratio & Ours / Voss & Ratio \\
\midrule
$[[84,6,10]]$ wt-6  & $[[48,4,6]]$ wt-6  & 7.10 / 6.14 & $1.16\times$ & 7.14 / 3.00 & $2.38\times$ \\
$[[140,6,14]]$ wt-6 & $[[96,4,8]]$ wt-5   & 8.02 / 7.89 & $1.02\times$ & 8.40 / 2.67 & $3.15\times$ \\
$[[196,6,12]]$ wt-6 & $[[96,4,8]]$ wt-5   & 8.49 / 7.89 & $1.08\times$ & 4.41 / 2.67 & $1.65\times$ \\
$[[54,8,6]]$ wt-6   & $[[48,4,6]]$ wt-6  & 3.46 / 6.14 & $0.56\times$ & 5.33 / 3.00 & $1.78\times$ \\
$[[54,14,5]]$ wt-8  & $[[30,4,5]]$ wt-5   & 2.57 / 4.51 & $0.57\times$ & 6.48 / 3.33 & $1.94\times$ \\
$[[128,20,8]]$ wt-8 & $[[96,4,8]]$ wt-5   & 5.05 / 7.89 & $0.64\times$ & 10.00 / 2.67 & $3.75\times$ \\
\bottomrule
\end{tabular}
\end{table}

	Among weight-6 codes, the three $k = 6$ codes achieve $2.0$--$2.4\times$ the pseudothreshold of BB~$[[72,12,6]]$ and up to $1.4\times$ the $kd^2/n$ ratio. The $[[54,8,6]]$ code is comparable to BB~$[[72,12,6]]$ in pseudothreshold ($0.98\times$). Against BB~$[[144,12,12]]$, the $[[140,6,14]]$ code achieves $1.11\times$ the pseudothreshold. Compared with Voss et al.\ codes of the same stabilizer weight, $[[84,6,10]]$ exceeds Voss et al.~$[[48,4,6]]$ by $1.16\times$ in $p_0$ and $2.38\times$ in $kd^2/n$.

	The weight-8 trivariate bicycle codes trade pseudothreshold for encoding rate and $kd^2/n$. Against Voss et al.\ codes, the $p_0$ ratios fall below~1 ($0.57$--$0.64\times$) while $kd^2/n$ ratios range from $1.94\times$ to $3.75\times$.

\subsection{Circuit-level depolarizing comparison}

	Table~\ref{tab:vs-bb-depol} compares extrapolated logical error rates from Table~\ref{tab:depol-results}. At $p = 10^{-3}$, the $[[140,6,14]]$ code achieves $p_L = 4 \times 10^{-8}$ and the $[[84,6,10]]$ code achieves $p_L = 2 \times 10^{-6}$, with pseudothresholds of $0.59\%$ and $0.53\%$ respectively. The high-distance codes benefit from steeper $p^{d/2}$ scaling ($d = 14$, $12$, and $10$) relative to BB $[[72,12,6]]$ ($d = 6$). These codes encode $k = 6$ logical qubits, compared to $k = 12$ for both BB codes; the lower $k$ allows the third cyclic dimension to be allocated toward higher distance.

\begin{table}[htbp]
\centering
\caption{Extrapolated per-round logical error rates from Table~\ref{tab:depol-results}. Ratios give factor by which our code's $p_L$ is lower than the BB code's; values ${<}\,1$ indicate our code has higher $p_L$. BB pseudothresholds are from~\cite{bravyi2024high} (Table~1); BB $p_L$ values are from our Monte Carlo data.}
\label{tab:vs-bb-depol}
\small
\begin{tabular}{lccccc}
\toprule
Our code & $p_L(10^{-3})$ & \multicolumn{2}{c}{vs BB $[[72,12,6]]$} & \multicolumn{2}{c}{vs BB $[[144,12,12]]$} \\
\cmidrule(lr){3-4}\cmidrule(lr){5-6}
 & & $p_L$ ratio & $p_0$ ratio & $p_L$ ratio & $p_0$ ratio \\
\midrule
$[[140,6,14]]$ & $4 \times 10^{-8}$ & $1{,}600\times$ & $1.23\times$ & $3.6\times$ & $0.91\times$ \\
$[[196,6,12]]$ & $7 \times 10^{-8}$ & $960\times$ & $1.21\times$ & $2.2\times$ & $0.89\times$ \\
$[[84,6,10]]$  & $2 \times 10^{-6}$ & $31\times$ & $1.10\times$ & $0.07\times$ & $0.82\times$ \\
$[[54,8,6]]$   & $1 \times 10^{-4}$ & $0.5\times$ & $1.00\times$ & $0.001\times$ & $0.74\times$ \\
\bottomrule
\end{tabular}
\end{table}

\section{Conclusion}
\label{sec:conclusion}

	We presented six ITB-QLDPC codes built from three cyclic shift matrices. The third dimension opens a design space beyond the bivariate bicycle codes of~\cite{bravyi2024high}.

	The best asymmetric codes achieve $kd^2/n$ up to $8.40$. In the code-capacity setting, the $[[140,6,14]]$ code achieves a pseudothreshold of $8.0\%$. Circuit-level depolarizing simulations give pseudothresholds of $0.53\%$ for $[[84,6,10]]$ and $0.59\%$ for $[[140,6,14]]$, with extrapolated logical error rates of $2 \times 10^{-6}$ and $4 \times 10^{-8}$ at $p = 10^{-3}$. On the SI1000 noise model with the Tesseract decoder~\cite{lee2025tesseract}, the $[[140,6,14]]$ code achieves $R_{\mathrm{rnd}} = 5.6 \times 10^{-5}$ at $p = 0.20\%$. The qubit connectivity is comparable to the bivariate bicycle codes and is compatible with architectures that support moderate-degree, non-planar qubit graphs, such as reconfigurable atom arrays and modular ion-trap networks.

	Thus, the third cyclic dimension offers a larger design space than the bivariate construction, and the codes presented here are competitive with the bivariate bicycle codes of~\cite{bravyi2024high} across all noise models tested.

\subsection{Future directions}

	The empirical condition $7 \mid \text{lcm}(\ell_1, \ell_2, \ell_3)$ for the $k = 6$ code set suggests a large design space: tori such as $(2, 7, 9)$, $(2, 7, 11)$, and $(3, 7, 7)$ may yield codes with higher distances. For codes where ILP branch-and-cut becomes impractical, the randomized algorithm in QDistRnd can provide distance upper bounds. Other open directions include bias-tailoring, designing optimized syndrome extraction circuits (potentially tailored to different hardware models, e.g.\ ion-chain or atom-array), and sliding-window decoding for continuous operation.

\section*{Acknowledgments}

We thank Ming Wang for a detailed review and many constructive suggestions, Balint Pato for a thorough reading and feedback, and Sahil Khan and Amin Idelhaj for helpful discussion.

\bibliographystyle{plain}

\end{document}